\documentclass[11pt,a4paper]{article}
\usepackage{hyperref}
\usepackage{graphicx}
\usepackage{amsmath}
\providecommand{\keywords}[1]
{
  \small
  \textbf{\textit{Keywords:}} #1
}
\begin{document}
%
\title{\bf Time to critical condition in emergency services\footnote{
Math. Comput. Appl. {\bf 26}~(4), 70 (2021).
DOI: \href{https://doi.org/10.3390/mca26040070}{10.3390/mca26040070}}
}

\author{Pedro A.\ Pury\footnote{ORCID
\href{https://orcid.org/0000-0003-2229-0590}{0000-0003-2229-0590}
}
\\
Facultad de Matem\'atica, Astronom\'\i a,
F\'\i sica y Computaci\'on \\
Universidad Nacional de C\'ordoba \\
Ciudad Universitaria, X5000HUA C\'ordoba, Argentina \\
\texttt{pedro.pury@unc.edu.ar}}

\date{October 15, 2021}
%
\maketitle
%
\begin{abstract}
Providing uninterrupted response service is of paramount importance
for emergency medical services, regardless of the operating scenario.
Thus, reliable estimates of the time to the critical condition,
under which there will be no available servers to respond
the next incoming call, become very useful measures
of the system's performance.
In this contribution, we develop a key performance indicator
by providing an explicit formula for the average time
to the shortage condition. Our analytical expression
for this average time is a function of the number of parallel
servers and the inter-arrival and service times.
We assume exponential distributions of times in our analytical
expression but for evaluating the mean first-passage time
to the critical condition under more realistic scenarios
we validate our result through exhaustive
simulations with lognormal service time distributions.
For this task we have implemented an simulator in $R$.
Our results indicate that our analytical formula
is an acceptable approximation under any situation
of practical interest.
\end{abstract}
%


\section{Introduction}
\label{sec:intro}

The problem of assigning resources to respond to a stochastic
demand is a ubiquitous topic in operational research.
The trade-off between service quality and operational efficiency
is a crucial aspect of the Emergency Medical Services (EMS), where
the lives of patients depend on the timeliness of care.
Thus, the development of Key Performance Indicators (KPIs)
to objectively quantify the performance across the operational,
clinical and financial departments is a current demand
of an industry increasingly data driven. KPIs are the basic tools
for planners and the nature of each KPI selects a particular feature
of the system and determines its data gathering strategy.

Among the most intuitive and used operational KPIs in EMS
are the successive times involved in the service cycle:
call reception, patient triage, dispatch, ambulance turnout,
travel from the base to the emergency site, paramedic care,
eventual transfer of the patient to a hospital,
and return of the ambulance to its base.
Response time (the interval between the reception of an emergency
call and the arrival of a paramedic at the scene of the event)
is a common operational metric of EMS and it is considered a good
indication of the quality offered by the service~\cite{SSM+18}.
One reason for its popularity as KPI resides in the fact
that it is directly quantifiable and easily understood
by the public and policy makers. Additionally, the EMS industry
has the goal of providing care within eight minutes for cardiac
arrest~\cite{EBH79} and major trauma~\cite{FHSI95}.
However, there is evidence that exceeding that response time
criterion does not affect patient survival after a traumatic
impact injury~\cite{PM02,MMS+13}.
Moreover, solutions that only focus on shortening the response
time are cost prohibitive and put the safety of patients,
attendant crew and the public at risk~\cite{ZS10}.
A rational approach to the ambulance business process should
simultaneously consider multiple metrics and operational
trade-off between administrator-oriented and patient-centric
KPIs~\cite{SD08}.

One of the most important aspects of emergency medical
management is avoiding the oversaturation of the system.
Therefore, in this work we consider the First-Passage Time
(FPT)~\cite{Win03} to the critical condition, under which there
will be not available servers to respond the next incoming call.
Criticality prediction is of special interest for the
quality of the medical service (response time off-target)
as well as for the financial management of the service,
given that, as we will see, the critical condition strongly
depends on the number $L$ of ambulances simultaneously
in service and because queueing a call may involve transferring
it to another EMS.
Any system operating under fixed conditions with a given number
of servers and  First-Come First-Served (FCFS) discipline
is a discrete one-dimensional stochastic process over
the occupation states of servers.
Therefore, the first-passage time to the state of oversaturation,
in which there are no available servers, is finite.
The question is how long is that time. Thus, the mean first-passage
time (MFPT) becomes a relevant key performance indicator for
operational condition in EMS.

In urban emergency services logistics there are two distinct fields:
capacity planning and location analysis.
Both fields are interrelated in the districting problem
or how the region should be partitioned into areas of primary
responsibility (districts)~\cite{Lar74,TWM07}.
Here, MFPT can be applied to an operational subarea or district,
preferably intended to be independently served by a subset
of ambulances (intradistrict dispatches).
In this case, MFPT gives us the average time to request an ambulance
from another operational zone to answer the next emergency call
when the primary equipment is busy (interdistrict dispatches).
MFPT can also be an useful KPI in the decision-making process
of emergency departments~\cite{UZMN17,CIM19} where MFPT provides
the average time to a shortage of intensive therapy beds when FCFS
discipline is used after the patient's triage.

Queueing theory has been widely applied in health care
in the last 70 years~\cite{LS13}.
Since Larson's seminal article~\cite{Lar74}, quite a few
queueing models have been developed to incorporate
the intrinsic probabilistic nature of urban EMS,
derived from the Poisson nature of the call arrival
process and the variability in service times.
Multiple queueing systems have been developed that respond
with different emphasis to the KPIs selected in each case.
In this contribution we use a birth--death process to properly
analyse the dependence of MFPT on the number of servers and the
rest of system's operational parameters.
The birth--death process is basic to queueing models involving
exponential inter-arrival and service times distributions.
In the Kendall-Lee notation we have
$M/M/L/FCFS/\infty/\infty$~\cite{Win03}.
Thus, our analytical model is based only on two average times:
$T_C$, the mean inter-arrival time and
$T_S$, the mean service time of a single server, that is,
the time it takes for an ambulance to complete a trip,
from the instant a call is assigned until the release of this server.
Several analytical results are well known in operational research
under that assumption~\cite{Win03}.

However, experimental evidence from an emergency service indicates
that service time distributions are well fitted by lognormal
distributions~\cite{BGM+05,IBE08,BKMB15,CIM19}.
Hence, our objective is also to numerically evaluate the
deviation between analytical and simulated results for MFPT.
Thus, we present an $R$--simulator for a system of $L$
servers in parallel with general distributions for
inter-arrival and service times and FCFS discipline:
$GI/G/L/FCFS$~\cite{Win03}.
This tool allows the user to calculate the key performance
indicators of direct interest in the industry beyond the known
analytical results, limited to exponential distributions.
Particularly, we show our work on Mean First-Passage Time
(MFPT) to system critical condition.

In this way, the motivation of our work is two-fold.
On the one hand, we provide an explicit analytical expression
for the MFPT to the critical condition, and, on the other hand,
under more realistic conditions, we analyze the validity of
our assumptions through exhaustive simulations.
In Section II we provide our analytical expression for MFPT
and explore the generic nature of the method, postponing
detailed mathematical derivations to the appendices.
Also in that section we describe the simulation framework
for experimentation.
Section III deals with the numerical results and last,
in section IV, we discuss the importance of our contribution.

\section{Model and Simulator}
\label{sec:MM}

In this section we develop an analytical closed--form solution
for the MFPT and present a simulation framework based on discrete
events.

\subsection{Markov chain model for servers in parallel}
\label{sec:model}

We consider a stochastic continuous-time birth-death
process~\cite{Win03} that describes the time evolution
of the occupation state of a set of $L$ servers in parallel.
Changes in the state of the Markov chain imply the release of
a server or putting one into action (if at least one is available).
The state with occupation $n$ corresponds to $n$ received calls
not completely served yet.
Thus, when $n=0$ all the servers are free and there are neither
trips in process nor calls in queue.
For $0 < n \leq L$, there are no waiting calls and $n$ servers
are in course of action. In an equivalent way, we can say that
$n$ calls are simultaneously being served.
Particularly, when $n=L$ the system is saturated, that is,
all servers are occupied. Even though there is not any call
in the waiting queue, all servers have been assigned to calls,
and consequently, there are not any server available
to process the next eventual incoming call.
For $n>L$, the system is oversaturated, and there are $n-L$ calls
in the waiting queue.
At any time, the system can change its state of occupation
between its nearest neighbors. Therefore, we denote the transition
rate from the state $n$ to $n+1$ by $\omega^+_n$, whereas
the transition rate toward the lower occupation state
($n \rightarrow n-1$) is $\omega^-_n$.
We assume that the time between calls is an exponential random
variable with the mean number of calls per unit time
$\lambda = 1/T_C$.
On the other hand, the service time is also an exponential variable
with the rate or mean number of services per unit time and per server
$\mu = 1 / T_S$.
Thus, random call arrival times and service times consumed in each
trip are generated from continuous time distributions.
$T_C$ and $T_S$ are the average inter-arrival and service time,
respectively.
In our particular problem with $L$ servers, the transition
probabilities rates of the birth-death process are defined
by~\cite{CS61}
%
\begin{equation}
\begin{array}{l}
\omega^+_n = \lambda \;\;\forall n \,, \\
\omega^-_n =
\left\{
\begin{array}{rl}
n \, \mu  & \mbox{for } n \leq L \,, \\
L \, \mu  & \mbox{for } n \geq L \,.
\end{array}
\right.
\rule{0cm}{0.7cm}
\end{array}
\label{Lservers}
\end{equation}
In this manner, $\omega^+_n$ results constant and only $\omega^-_n$
depends on the state of the system.
Then, all the experimental information needed to characterize
our theoretical model are the average times $T_C$ and $T_S$.

For fixed values of $T_C$ and $T_S$, the Markov process always
reach the state $L+1$, that is, the critical condition in which
the incoming call could not be served.
Therefore, we focus our interest in the first-passage time (FPT)
to the state $L+1$. That is the time needed by the system to reach
the critical situation (first call is derived to the waiting queue)
given an initial state without queue ($0 \leq n \leq L$).
Following previous experience~\cite{PC03}, the MFPT, $T(n)$,
from the initial state $n = 0, \dots, L$, can be
written as
\begin{equation}
\begin{array}{lcl}
T(0) &=&
T_C \left(
L+1 + \displaystyle\sum_{k=0}^{L-1} \,
\frac{\gamma^{-k}}{k!} \sum_{i=k+1}^{L} i! \,\gamma^i
\right) \;,
\\
T(1) &=& T(0) - T_C \;,
\\
T(n) &=& T(0) - T_C \left(
n + \displaystyle\sum_{k=0}^{n-2} \,
\frac{\gamma^{-k}}{k!} \sum_{i=k+1}^{n-1} i! \,\gamma^i
\right)
\;\; \mbox{for } 2 \leq n \leq L\,,
\label{MFPT:critical}
\end{array}
\end{equation}
where the parameter $\gamma = \mu/\lambda = T_C/T_S$
is the inverse of Erlang's rate~\cite{Win03}.
The derivation and mathematical details of Eq.~(\ref{MFPT:critical})
are worked out in Ref.~\cite{PC03} and Appendix~\ref{sec:MFPT}.
In this way, knowing $\gamma$ and $T_C$, the expressions
of Eq.~(\ref{MFPT:critical}) can be numerically evaluated
in a very direct way.

The average involved at this stage is over realizations
of the stochastic process.
Under actual operating conditions, where the dispatcher knows
the system stress in real-time, Eq.~(\ref{MFPT:critical}) makes
it possible to predict the MFPT to respond accordingly.
However, if we want to predict the MFPT under prospective stress
conditions, we request for a quantity independent of the initial
state in order to define a performance measure.
Therefore, we need an average over $n = 0, \ldots, L$.
For this purpose, we define
\begin{equation}
<T> = \sum_{n=0}^L \, P(n) \,T(n) \,,
\label{<T>}
\end{equation}
where $P(n)$ is the probability of residence in the state $n$.
To perform this calculation, we need to know the conditional
probability of being at state $n$ at time $t$ given the initial
state $m$, $P(n|m)(t)$. In order to simplify the problem,
we propose to calculate $P(n)$ in the steady state regime,
which is independent of the initial condition~\cite{Win03}:
$P(n) = \lim_{t \rightarrow \infty} P(n|m)(t)$.
Moreover, given that we are interested in the FPT to the critical
condition (first jump to state $L+1$), we will approximate $P(n)$
by working with the finite Markov chain with reflecting boundaries
at sites $0$ and $L$.
It is important to note that the steady state probabilities
of the finite chain are approximately the same as the residence
probabilities of our unbounded chain, since in the lapse before
FPT there are no calls in the queue.
Under these assumptions we obtain
\begin{equation}
P(n) = \displaystyle\frac{1}{S} \,\frac{\gamma^{-n}}{n!}\,,
\;\; \mbox{for } 0 \leq n \leq L\,,
\label{Pn}
\end{equation}
where $S$ is given by the normalization condition,
$S = \displaystyle\sum_{n=0}^{L} \,\frac{\gamma^{-n}}{n!}$.
The mathematical derivation of these expressions is relegated
to the Appendix~\ref{sec:SS}.
The truncated Poisson distribution given in Eq.~(\ref{Pn})
is known as Erlang $B$-formula and it has been proved that
this equilibrium distribution of the number of occupied
servers is independent of the form of the service time
distribution~\cite{Tak69}.
Moreover, the Erlang $B$-formula is also valid for heterogeneous
servers, provided however, that all servers have equal mean
service times $T_S$~\cite{Fak80}.

Eq.~(\ref{<T>}) plus Eqs.~(\ref{MFPT:critical}) and ~(\ref{Pn})
provide us a closed form expression for calculating the MFPT
averaged on initial states.

\subsection{Simulation framework}
\label{sec:simul}

To compare and analyze the prediction of Eqs.~(\ref{MFPT:critical})
and~(\ref{Pn}) with realistic situations, we have developed
a flexible discrete-event simulator (DES)~\cite{BCNN14}
with a process-oriented approach, that implements several
features inherent to EMS management.

\begin{figure}[!ht]
\begin{center}
\includegraphics[width=0.70\textwidth]{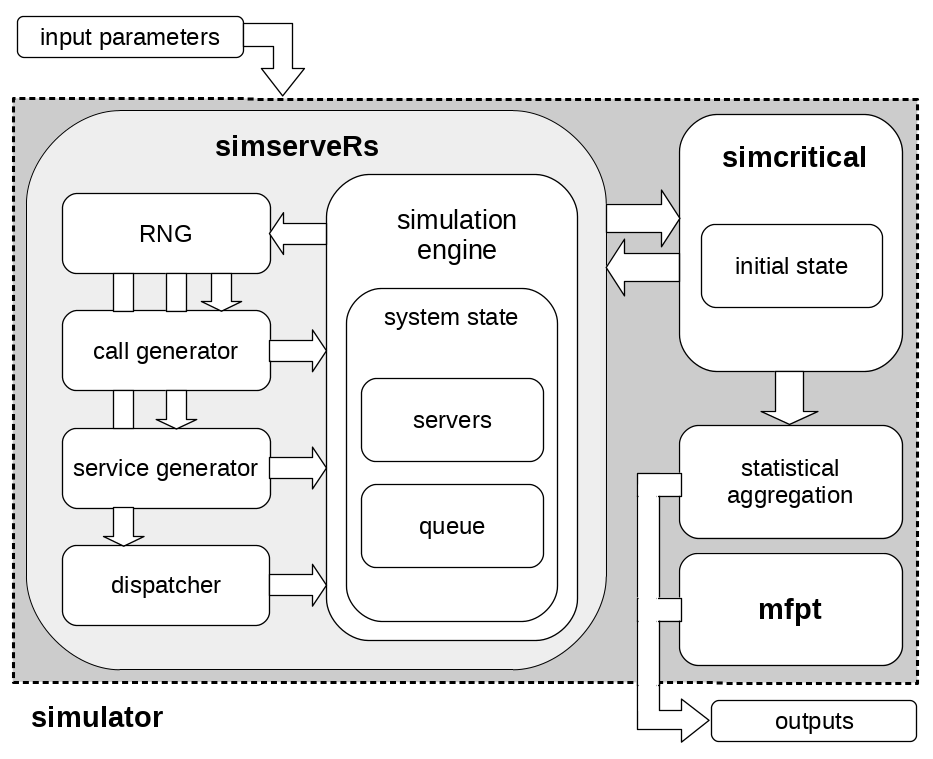}
\caption{Architecture scheme of the discrete-event simulator.}
\label{simulator}
\end{center}
\end{figure}

The architecture  of our simulator is outlined
in Figure~\ref{simulator}.
The input parameters configure the statistical distributions
and set the number of servers ($L$).
The proposed simulator consists of three main modules.
The first module, called {\tt simserveRs}, is the simulator kernel.
Each thread of simulation is triggered with a new call arrival.
Using a pseudo-random number generator (RNG), it draws a call time,
summing an exponential random value to the time of the previous call,
and compares it with the release times of busy servers.
Then, the kernel puts iteratively in service the older calls
in queue, whereas the release times are less than
the last call time (FCFS discipline)~\cite{Kri94}.
Afterward, if there are no available servers, the incoming call
is derived to the waiting queue; otherwise, the call is dispatched
to a server, picking it at random among the free ones.
At last, {\tt simserveRs} assigns a service time drawn from
the desired distribution.
Thus, at each event, the simulation engine updates the system state
composed by the servers and the queue.

The module {\tt simcritical} launches {\tt simserveRs} with the
wanted initial condition and stops the execution when the first
call is derived to the queue. Then, the average given by
Eq~(\ref{<T>}) is calculated and the aggregation over simulations
is performed.
The last module reckons the MFPT from each initial condition
using the analytical expressions given by Eq.~(\ref{MFPT:critical}).

The software is implemented in $R$ and it is available as open
source (see details in Ref.~\cite{simserveRs}).
The simulator can be simply adapted to incorporate most
of the features of an actual EMS operator, like disaggregate
the service time in its components (e.g. preparation of ambulance
at base, transit time, attention time and transit time to hospital)
and implementing dispatching policies with distance or traffic time
criteria.
Additionally, the extensions of the simulator to any kind
of distributions for inter-arrival times (e.g., Erlang)
and service times (e.g., gamma or lognormal) are direct.
Our simulator was developed in the second half of 2017
for a project related to process optimization in EMS management.
The comparison between our simulation outputs and the historical
data from an EMS operator showed a satisfactory statistical
agreement in several operating scenarios.
Over time, several generic DES frameworks for queueing systems
have been developed which delivers such functionalities
and implements more efficient methods (see Ref.~\cite{USA19}
and references therein).
However, for the practical reason of evaluating the results
of Section~\ref{sec:model}, the open source version of
our simulator becomes an appropriate tool.

\section{Results}
\label{sec:RR}

\begin{figure}[!ht]
\begin{center}
\includegraphics[width=0.70\textwidth]{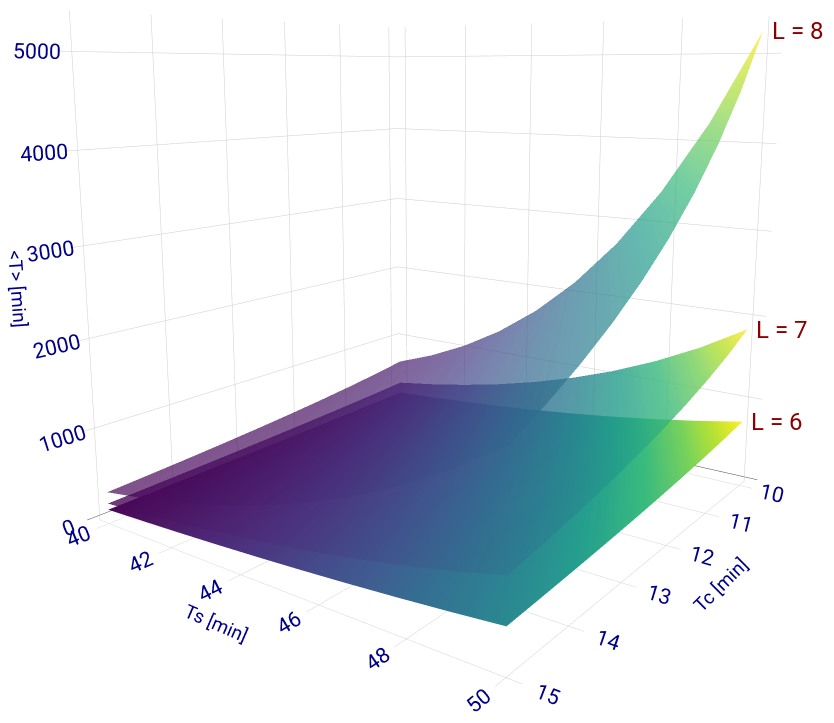}
\caption{Average MFPT as function of $T_C$ and $T_S$
for three values of $L$.}
\label{Fig.MFPT}
\end{center}
\end{figure}

In Figure~\ref{Fig.MFPT} we show the non-linear behavior
of $<T>$ as function of $T_C$ and $T_S$ and its strong
dependence on $L$, as they are derived from
Eqs.(\ref{MFPT:critical}--\ref{Pn}).
The non-linear nature of $<T>$ is given by the powers of $\gamma$
in Eqs.~(\ref{MFPT:critical}) and implies a sensitive dependence
of the time to the critical condition on its variables.
Thus, on the scale of the figure, variations in the order
of one minute in $T_C$ or $T_S$ may represent variations
of tens to thousands minutes in $<T>$.
The value of $L$ determines the number of terms
in Eqs.(\ref{MFPT:critical}) and (\ref{<T>}).
Therefore, adding or removing a server, with the same values
of $T_C$ and $T_S$, can make a substantial change in the value
of the average MFPT, as can be seen in Figure~\ref{Fig.MFPT}.
The sensitivity of our problem on its parameters is very difficult
to grasp intuitively and to predict from past experiences in practice.

The main reason for using birth and death queueing model ($M/M/L$)
is the fact that we can derive the closed-form result given
in Sec.~\ref{sec:MM}. However, the analytical prediction of any
performance measure need to the contrasted with numerical outputs
from more realistic scenarios.
As an illustration, we take values of inter-arrival and service
times measured by the EMS {\em Sistema de Urgencias del Rosafe}
(URG)~\cite{URG} in C\'ordoba, Argentina. URG is one of several
private EMS operators in a city of about 1.39 millions of inhabitants.
In 2016 URG operated a fleet of 9 ambulances that were usually
stationed in predefined parking spots scattered throughout
the metropolitan zone.
The operating scenario distinguishes several daily time bands within
which the mean values $T_C$ and $T_S$ are relatively constant,
although these, in turn, show seasonal changes throughout the year.

\begin{figure}[!hb]
\begin{center}
\includegraphics[clip,width=0.70\textwidth]{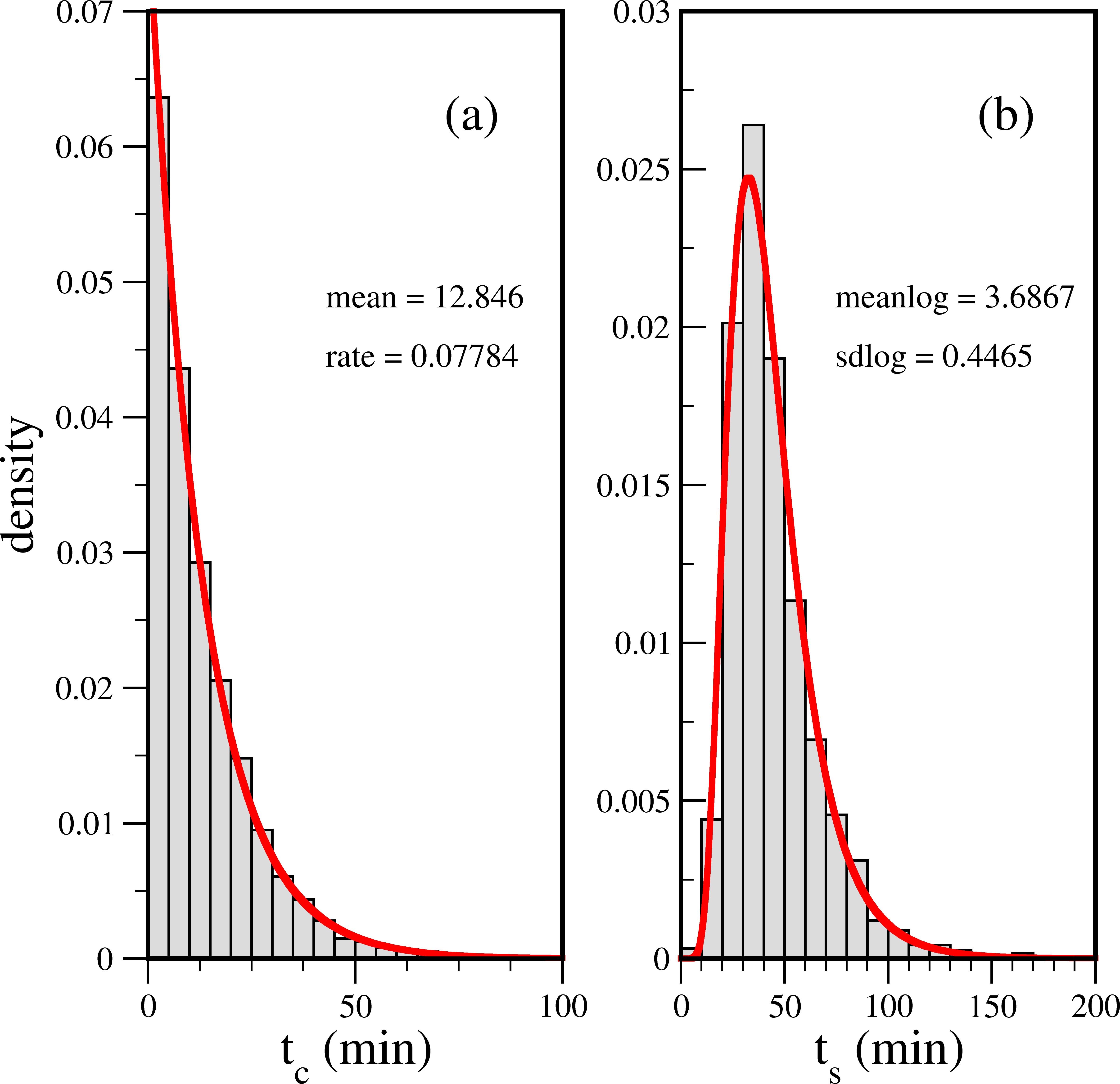}
\caption{Histograms of real data corresponding to 2568 calls:
(a) inter-arrival times, (b) service times.
Solid lines are the best fits:
(a) exponential, (b) lognormal.
The insets show the fitting parameters.}
\label{histosTcTs}
\end{center}
\end{figure}

In Figure~\ref{histosTcTs} we show histograms of real data
corresponding to 2568 calls received by URG between
May 1 and October 31, 2016, late evening (20:00 to 23:00 hours).
In this time period, we found good stationary statistics in
the data and no critical condition was reported in which there
were no ambulances available to serve a call.
The service time value measures the time elapsed from dispatch
until the ambulance is released. Very low service times correspond
to situations that were quickly resolved at scenes close to
the ambulance base, whereas the distribution tail involves complex
cases with the transfer of the patient to a hospital.
From the figure we can see that the inter-arrival times
fit perfectly with an exponential distribution
($T_C = 12.85\,$min, goodness-of-fit tests:
Kolmogorov-Smirnov $p$-value $= 0.73$,
Cramer-von Mises $p$-value $= 0.75$),
but the best fit for service times is with a lognormal distribution
($T_S = 44.1\,$min, goodness-of-fit tests:
Kolmogorov-Smirnov $p$-value $= 0.017$,
Cramer-von Mises $p$-value $= 0.052$).
Thus, in this case, it is apparent that the main limitation
of our analytical model is the assumption that the distribution
of service times is exponential.

The input traffic to a call centre is a nonstationary Poisson
process~\cite{Sze84,BGM+05,SD08}. However, the arrival rate
function $\lambda(t)$ is roughly constant over periods of few
hours~\cite{Hall71,BGM+05,BKMB15,UZMN17}.
Lognormal distributions for service times have already been
reported in the literature~\cite{BGM+05,IBE08,BKMB15,CIM19}.
The fact that the distribution of the sum of few independent,
but not necessarily identical, lognormal random variables,
could be approximated by a lognormal distribution~\cite{WMZ05},
may explain the experimental findings when we only use
two fitting parameters.

In Table~\ref{comparison} we show a comparison between
analytical results given by Eq.~(\ref{MFPT:critical})
and simulated MFPT from each initial state of a system with
seven servers to the critical condition.
The simulations were performed using the fitted distributions
in Figure~\ref{histosTcTs} and also using an exponential distribution
for service times with a mean value equal to that of the fitted
distribution in the right panel.
\begin{table}
\begin{center}
\begin{tabular}{c|rrr|rr}
\hline
\multicolumn{6}{c}{$T_C=5\,$min \rule{0pt}{12pt}} \\
\hline
  \multicolumn{1}{c|}{initial \rule{0pt}{12pt}} &
  \multicolumn{3}{c|}{exponential} &
  \multicolumn{2}{c }{lognormal} \\

state & analytic & simulated & {\large $\varepsilon$}$\%$ & simulated & {\large $\varepsilon$}$\%$ \\

0 & 67.6 & 67.2 $\pm$ 0.4 & 0,6 & 49.8 $\pm$ 0.3 & 35,7 \\
1 & 62.6 & 62.2 $\pm$ 0.4 & 0,5 & 44.8 $\pm$ 0.3 & 39,7 \\
2 & 57.1 & 56.7 $\pm$ 0.4 & 0,7 & 39.8 $\pm$ 0.3 & 43,5 \\
3 & 50.8 & 50.5 $\pm$ 0.4 & 0,6 & 34.7 $\pm$ 0.3 & 46,4 \\
4 & 43.7 & 43.4 $\pm$ 0.4 & 0,7 & 29.4 $\pm$ 0.2 & 48,6 \\
5 & 35.4 & 35.1 $\pm$ 0.3 & 0,9 & 23.7 $\pm$ 0.2 & 49,4 \\
6 & 25.8 & 25.4 $\pm$ 0.3 & 1,6 & 17.1 $\pm$ 0.2 & 50,9 \\
7 & 14.2 & 14.1 $\pm$ 0.2 & 0,7 &  9.3 $\pm$ 0.2 & 52,7 \\

\hline
\multicolumn{6}{c}{$T_C=10\,$min \rule{0pt}{12pt}} \\
\hline
  \multicolumn{1}{c|}{initial \rule{0pt}{12pt}} &
  \multicolumn{3}{c|}{exponential} &
  \multicolumn{2}{c }{lognormal} \\

state & analytic & simulated & {\large $\varepsilon$}$\%$ & simulated & {\large $\varepsilon$}$\%$ \\

0 & 315.2 & 310.9 $\pm$ 2.5 & 1,4 & 244.2 $\pm$ 2.1 & 29,1 \\
1 & 305.2 & 300.7 $\pm$ 2.5 & 1,5 & 234.2 $\pm$ 2.1 & 30,3 \\
2 & 292.9 & 288.4 $\pm$ 2.5 & 1,6 & 223.9 $\pm$ 2.1 & 30,8 \\
3 & 277.3 & 272.7 $\pm$ 2.5 & 1,7 & 212.0 $\pm$ 2.1 & 30,8 \\
4 & 256.7 & 252.3 $\pm$ 2.5 & 1,7 & 197.7 $\pm$ 2.1 & 29,8 \\
5 & 228.1 & 224.3 $\pm$ 2.5 & 1,7 & 176.9 $\pm$ 2.0 & 28,9 \\
6 & 185.6 & 182.1 $\pm$ 2.4 & 1,9 & 145.6 $\pm$ 2.0 & 27,5 \\
7 & 117.7 & 115.2 $\pm$ 2.0 & 2,2 &  92.1 $\pm$ 1.7 & 27,8 \\

\hline
\multicolumn{6}{c}{$T_C=15\,$min \rule{0pt}{12pt}} \\
\hline
  \multicolumn{1}{c|}{initial \rule{0pt}{12pt}} &
  \multicolumn{3}{c|}{exponential} &
  \multicolumn{2}{c }{lognormal} \\

state & analytic & simulated & {\large $\varepsilon$}$\%$ & simulated & {\large $\varepsilon$}$\%$ \\

0 & 1378.4 & 1380 $\pm$ 13 & 0,1 & 1201 $\pm$ 12 & 14,8 \\
1 & 1363.4 & 1365 $\pm$ 13 & 0,1 & 1186 $\pm$ 12 & 15,0 \\
2 & 1343.3 & 1345 $\pm$ 13 & 0,1 & 1169 $\pm$ 12 & 14,9 \\
3 & 1314.6 & 1317 $\pm$ 13 & 0,2 & 1147 $\pm$ 12 & 14,6 \\
4 & 1270.4 & 1272 $\pm$ 13 & 0,1 & 1114 $\pm$ 12 & 14,0 \\
5 & 1195.1 & 1198 $\pm$ 13 & 0,2 & 1056 $\pm$ 12 & 13,2 \\
6 & 1052.2 & 1058 $\pm$ 12 & 0,5 &  937 $\pm$ 12 & 12,3 \\
7 &  745.4 &  751 $\pm$ 12 & 0,7 &  670 $\pm$ 11 & 11,3 \\

\hline
\end{tabular}
\end{center}
%
\caption{Comparison among analytic and simulated values of MFPT
(in minutes) for a system with seven servers.
The simulated values are reported with its standard error for
$10000$ simulations.
In both cases, the mean service time is {$T_S=44.1\,$}min, but
whereas in the first case the probability distribution of service
time is exponential, in the second case is lognormal
with {\tt meanlog} $= 3.6867$ and {\tt sdlog} $= 0.4465$.
The percentage errors are that of the simulated values
with respect to the analytical predictions.}
\label{comparison}
\end{table}
The analytical predictions agree perfectly with the simulations
for exponentially distributed service times in all cases.
For lognormal distribution of service times, the deviations
between prediction and simulation are about of $50\%$
in the worst situations.
However, these cases are of little practical interest.
Given a fixed number of servers, low values of $T_C$
imply very low MFPT values incompatible with EMS response
times, while situations with very high values of MFPT
are often related with idle infrastructure,
which are also avoided in practice.

\begin{figure}[!ht]
\begin{center}
\includegraphics[clip,width=0.70\textwidth]{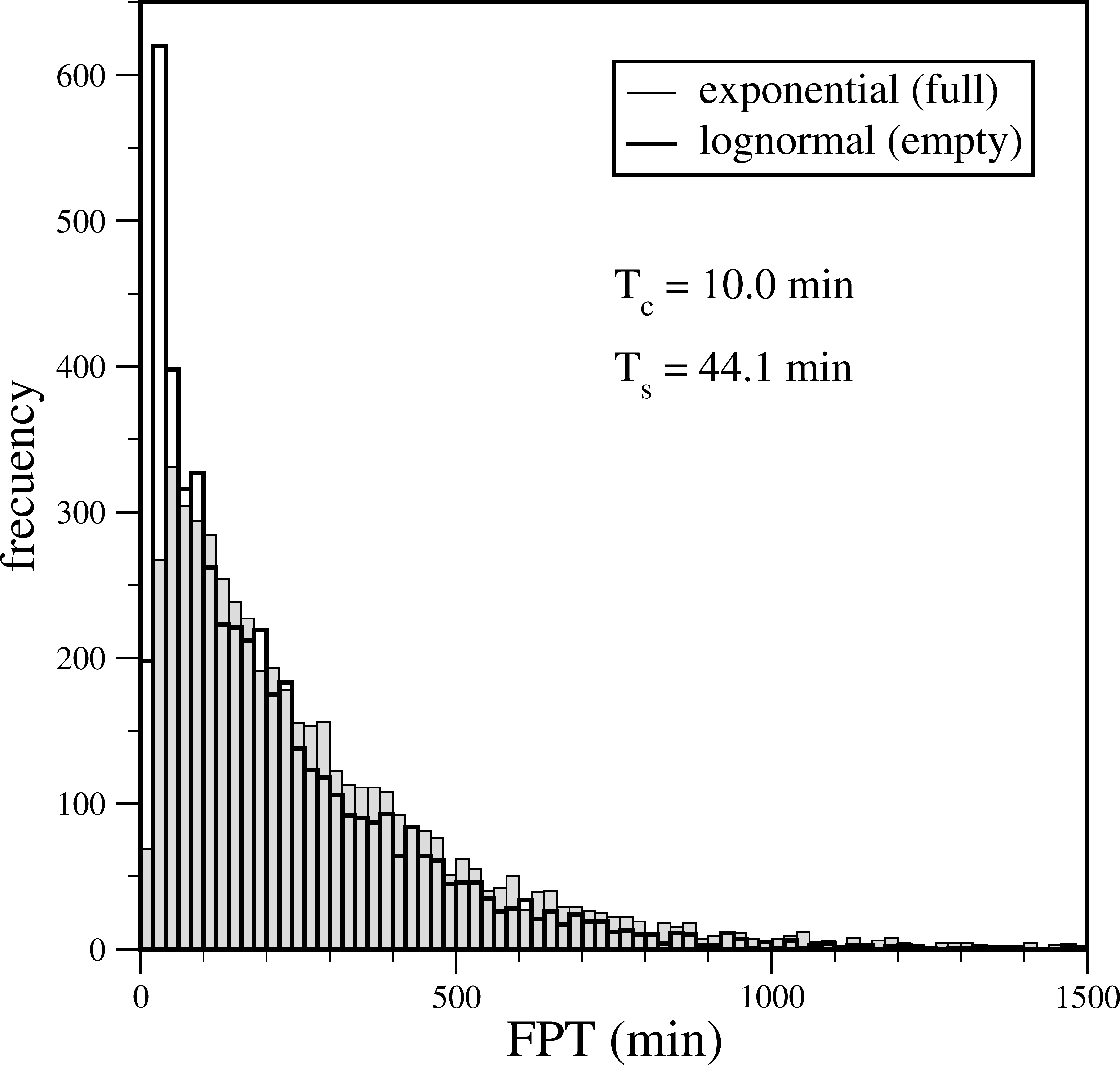}
\caption{Histograms of FPT to critical condition based on $5000$
simulations from an initial state with 4 of 7 servers occupied.
We compare two service time distributions with the same value
of $T_S$: exponential (parameter $= 1/T_S$) and lognormal
({\tt meanlog} $= 3.6867$ and {\tt sdlog} $= 0.4465$).}
\label{histos4}
\end{center}
\end{figure}

In order to study the differences in the values of FPT under
different service time distributions, in Figure~\ref{histos4}
we superimpose two histograms of simulated values for a system
with seven servers. Both cases correspond to an initial condition
with four occupied servers and the same exponential inter-arrival
time distribution, but we use two different service time
distributions with the same mean value.
The FPT distribution with exponential service times is broader
than the lognormal counterpart. Therefore, the MFPT under these
conditions is shorter for the lognormal service time distribution
(also see entry for initial state equal to 4 in Table~\ref{comparison}).

\begin{figure}[!ht]
\begin{center}
\includegraphics[width=0.40\textwidth]{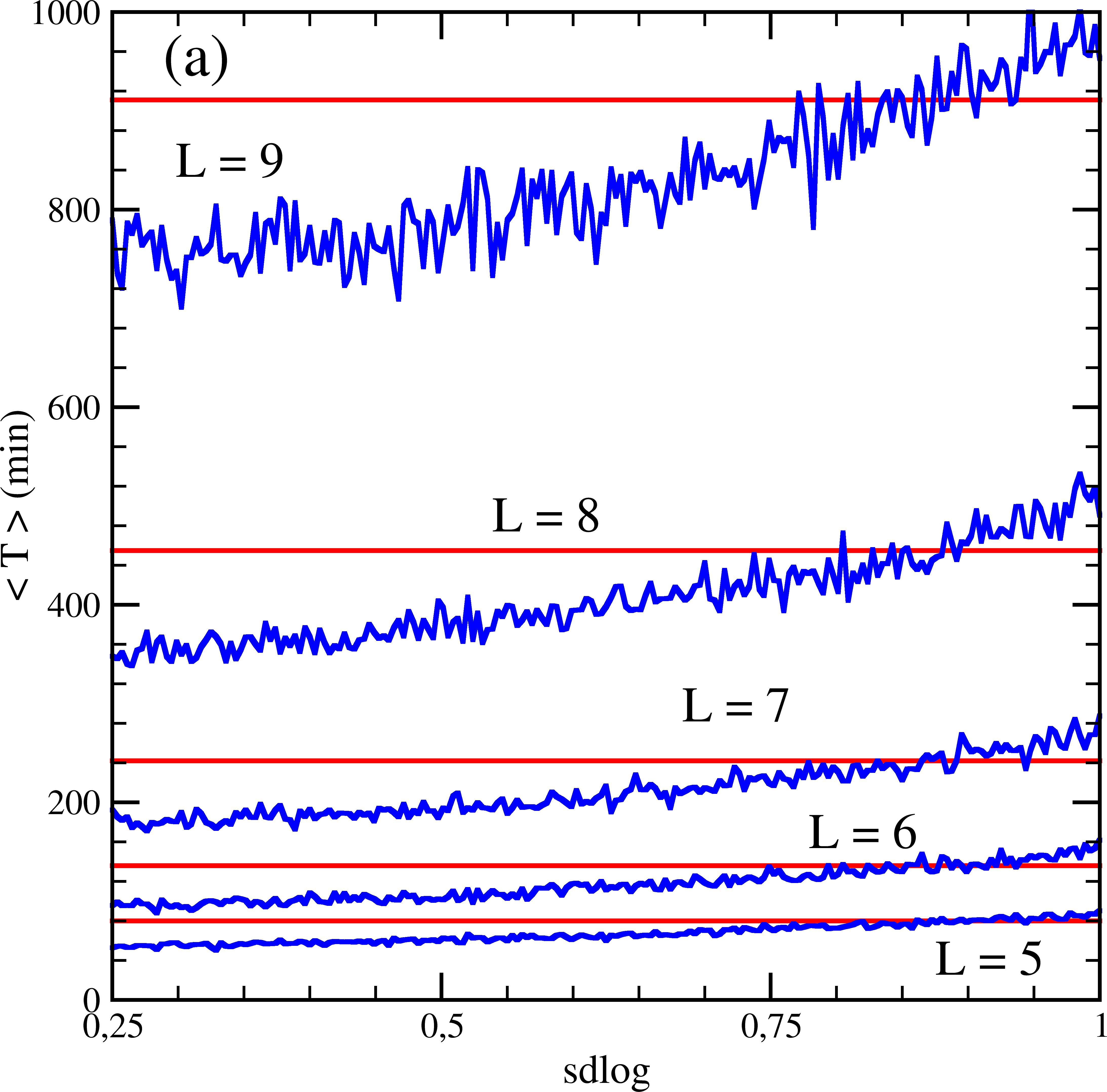}
\includegraphics[width=0.405\textwidth]{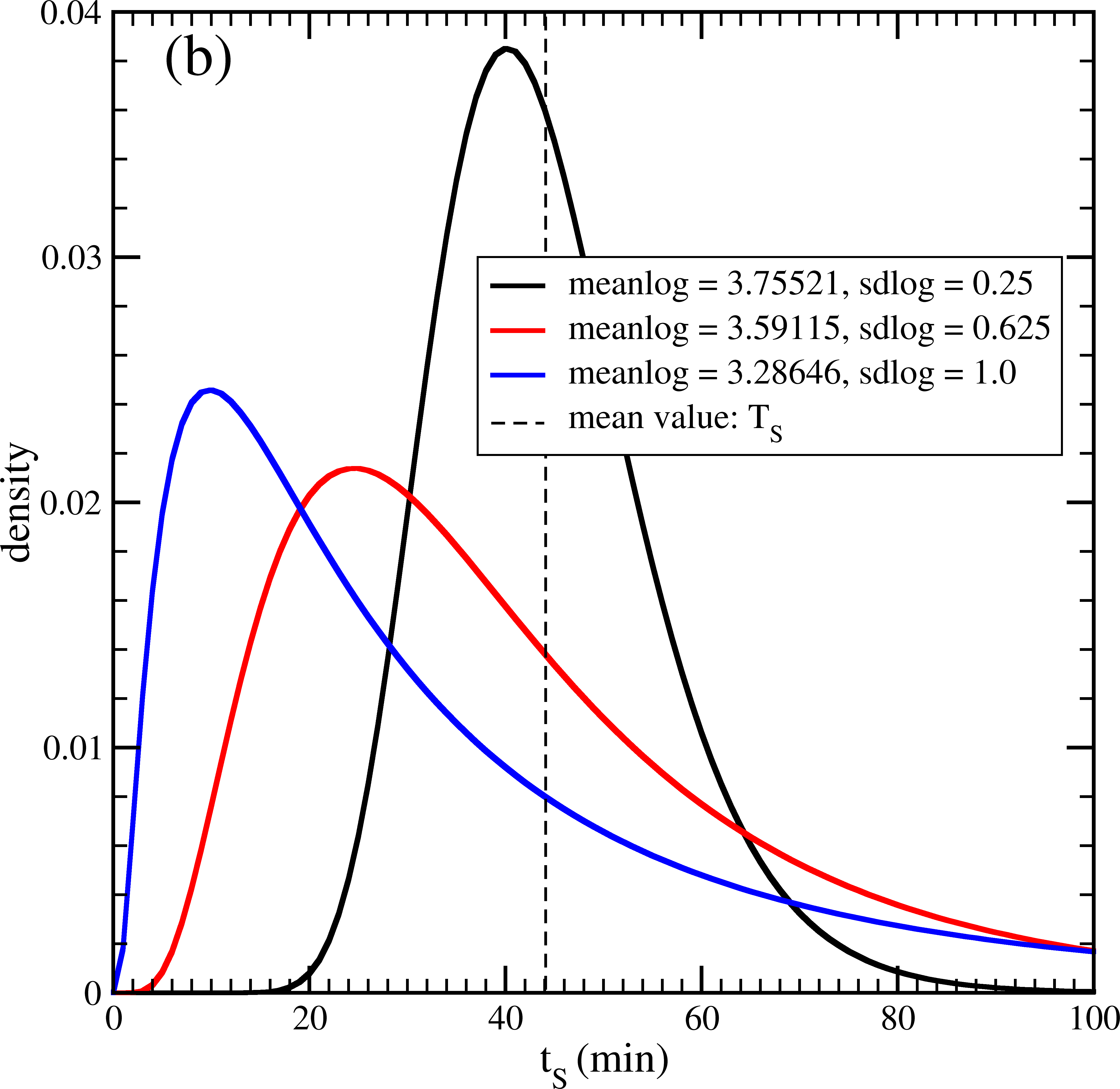}
\caption{
{\bf (a)} Analytic (red) and simulated (blue)
average MFPT with lognormal distributed  service times with
fixed  $T_S=44.1\,$min ($\mbox{\tt sdlog} \in [0.25,1]$ and
$\mbox{\tt meanlog} = \ln(T_S) - \mbox{\tt sdlog}^2/2$)
and $T_C=10\,$min. Simulated points for each value of {\tt sdlog}
are averages based on 1000 runs.
{\bf (b)} Sketches of lognormal densities corresponding
to the extreme values (black and blue) and to the middle
value (red) of {\tt sdlog} in panel (a).
}
\label{lognorm}
\end{center}
\end{figure}

Now we investigate what happens with the residence probabilities
in the long time limit. The analytical result given by
the Erlang formula has been proved for systems without
queue or $M/G/L$ blocking systems~\cite{Tak69,Fak80}.
That is, if every server is busy when a call arrives,
the call is lost; however, it is instructive to analyze
the situation of a system with queue.
The probability of residence in the queue, $P(n>L)$, is equivalent
to the probability absorbed in the site $L+1$ of the Markov chain.
From Eq.(\ref{pinfty}), we can see that the probabilities
in Eq.(\ref{Pn}) are the renormalized residence probabilities
of a system with a queue in the interval $\left[ 0,L \right]$.
To find out if this is also the case for lognormal service time
distributions, we run our simulator $10^7$ minutes in order to
visit each state several times.
In the first column of Table~\ref{residence}, we show the analytic
result of Eqs.~(\ref{pinfty}--\ref{piq}) and compare this prediction
with the simulated values using four different service time
distributions: an exponential and three different lognormal
service time distributions, all of these with the same value
$T_S = 44.1\,$min ($T_C = 10\,$min in all cases).
The value {\tt sdlog}$ = 0.25$ corresponds to the almost symmetric
case of the lognormal distribution, and {\tt sdlog}$ = 1$ corresponds
to the more asymmetric situation (see Figure~\ref{lognorm}b).
The exponential distribution is the case completely asymmetric,
where the distribution does not have a maximum value.
For comparison, we also introduce the middle value
{\tt sdlog}$ = 0.625$.
\begin{table}
\begin{center}
\begin{tabular}{c|ccccc}
\hline
state & analytic & exponential & lognorm(a) & lognorm(b) &  lognorm(c) \\
0 & 0.012 & 0.011 & 0.011 & 0.011 & 0.012 \\
1 & 0.051 & 0.050 & 0.049 & 0.050 & 0.052 \\
2 & 0.112 & 0.112 & 0.109 & 0.111 & 0.113 \\
3 & 0.165 & 0.165 & 0.164 & 0.164 & 0.165 \\
4 & 0.182 & 0.182 & 0.184 & 0.182 & 0.181 \\
5 & 0.160 & 0.160 & 0.167 & 0.163 & 0.158 \\
6 & 0.118 & 0.118 & 0.127 & 0.122 & 0.115 \\
7 & 0.074 & 0.074 & 0.085 & 0.081 & 0.075 \\
q & 0.126 & 0.127 & 0.104 & 0.114 & 0.130 \\
\hline
\end{tabular}
\end{center}
\caption{Comparison among analytic and simulated long-run
probabilities of residence in each state of a system with
seven servers, where q denotes the state with queued calls.
All simulations correspond with a simulated running time
of $1 \times 10^7\,$min.
In all cases, $T_C=10\,$min and $T_S=44.1\,$min, but the
parameters of the lognormal distributions are given
by the pairs  ({\tt meanlog}, {\tt sdlog}):
(a) $(3.75521,0.25)$, (b) $(3.59115,0.625)$, and (c) $(3.28646,1.0)$.
See Figure~\ref{lognorm}b.}
\label{residence}
\end{table}
The difference among analytical and simulated values are
less than $5\%$ in all states with exception of the queue.
The analysed case with queue is a worse situation than the
finite chain with both reflecting ends, used in the derivation
of Eq.~(\ref{Pn}), because of the probability of residence
in the queue may be greater than the saturated state.
%
Therefore, we find it is valid to use the analytic result
of Eq.~(\ref{Pn}) for taking the average in Eq.~(\ref{<T>}),
and also in the simulation of MFPT with lognormal service time
distributions in a system with queue.

\begin{figure}[!ht]
\begin{center}
\includegraphics[width=0.40\textwidth]{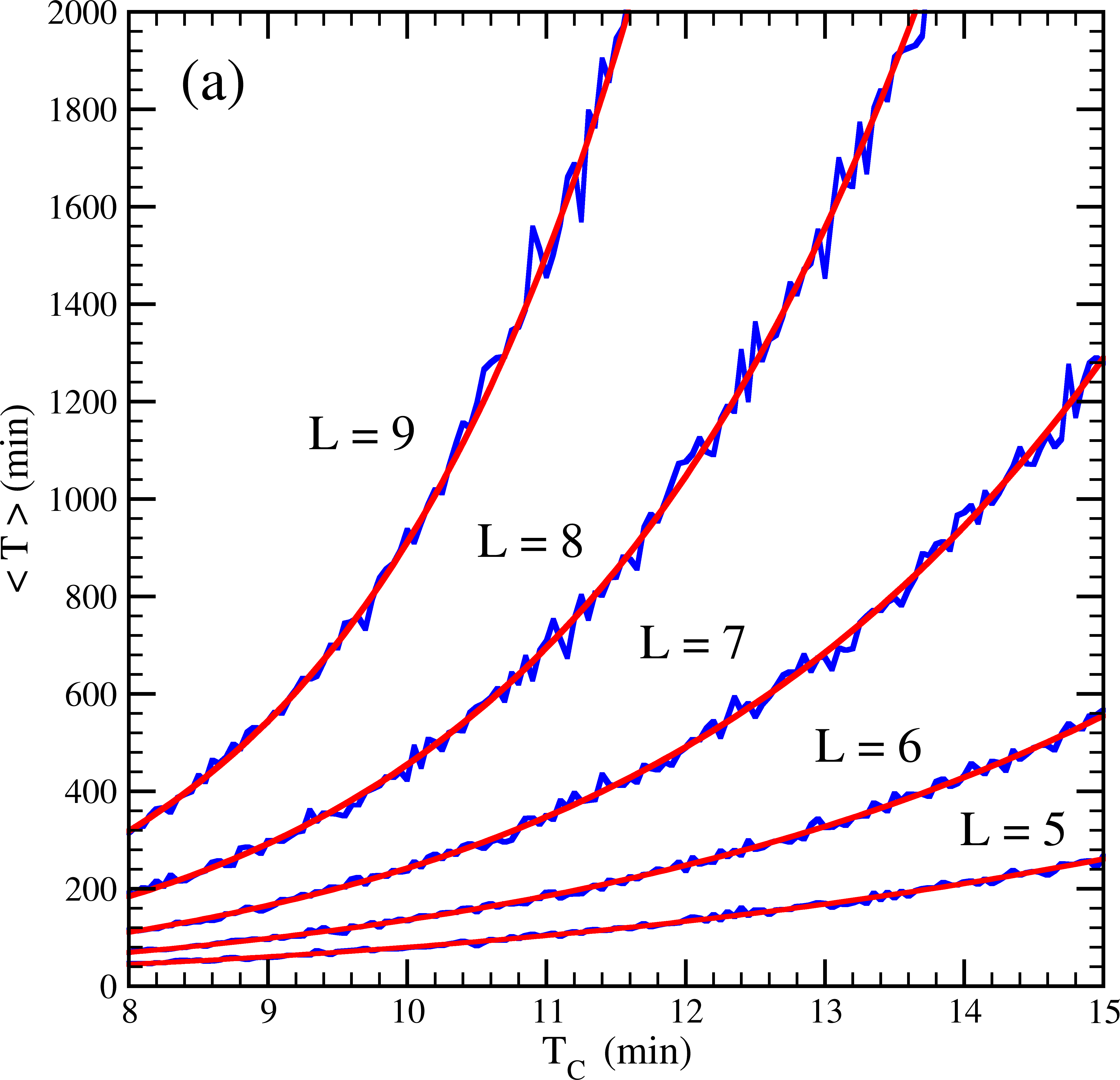}
\includegraphics[width=0.40\textwidth]{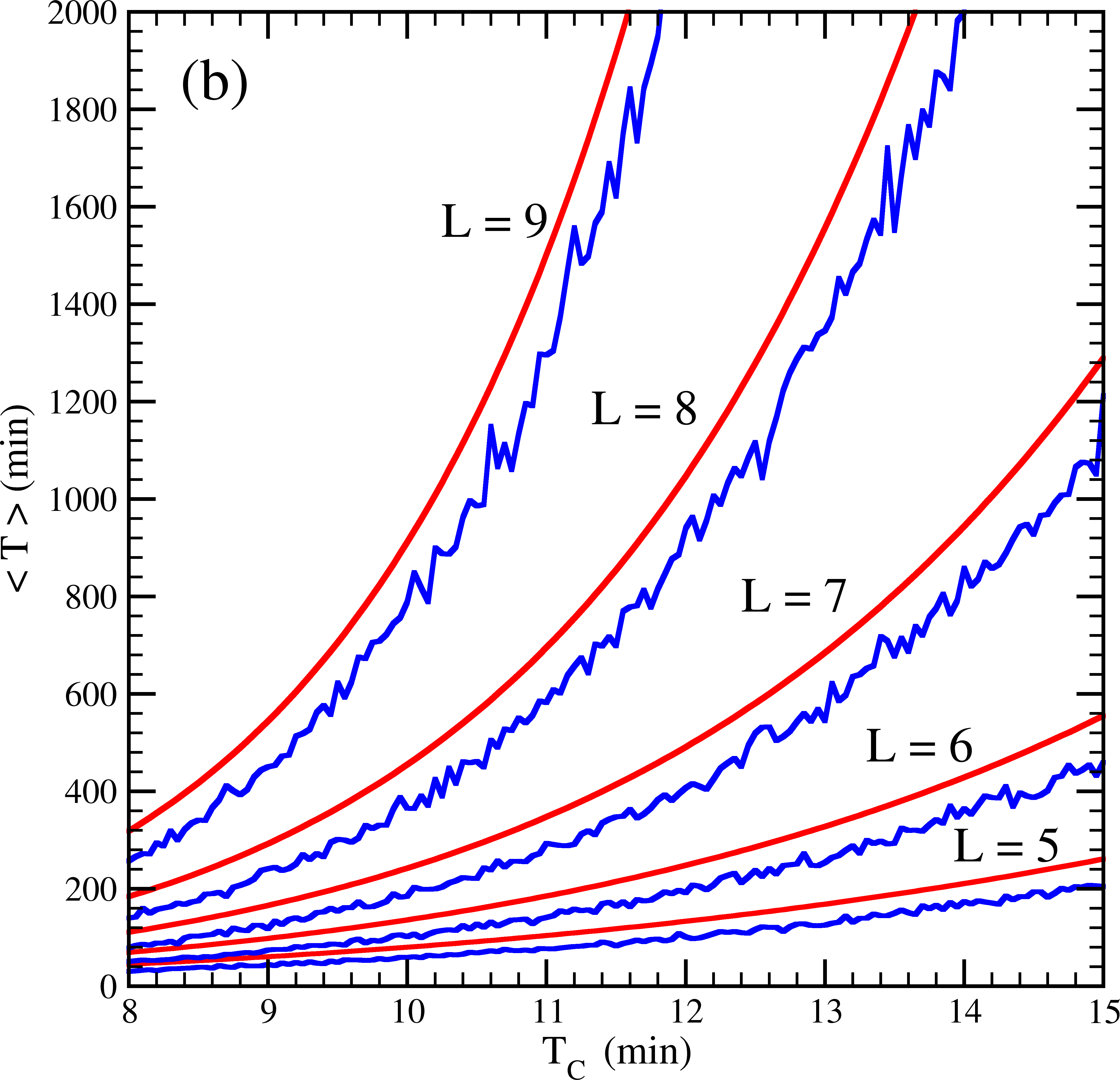}
\caption{Analytic (red) and simulated (blue)
average MFPT with {\bf (a)} exponential and {\bf (b)} lognormal
distributed  service times
({\tt meanlog} $= 3.6867$ and {\tt sdlog} $= 0.4465$).
In both cases, $T_S=44.1\,$min.
Simulated values for each value of $T_C$ correspond to an average
based on 1000 executions.}
\label{explogn}
\end{center}
\end{figure}

In Figure~\ref{explogn}, we show the plots of MFPT averaged over
the initial conditions, $<T>$, as function of the mean inter-arrival
time $T_C$, using the probabilities given by Eq.~(\ref{Pn}).
We have sketched a characteristic situation for the service time
and we considered several number of servers $L=5, \dots, 9$.
The curves clearly show the non-linear behavior of $<T>$
and allow us to evaluate the quality of the fit for the simulated
situations achieved with our analytical expression.
Again, in the left panel, we find excellent agreement among
the analytical predictions and the simulations for exponentially
distributed service times.
In contrast, in the right panel, for lognormal distributed service
times, we see that the analytical curves always run over
the corresponding simulation.
This fact has just been seen in Figure~\ref{histos4}, where
the FPT distribution has a longer tail in the case of service times
distributed exponentially with respect to lognormal service times.
Thus, the mean value of the FPT distribution (MFPT) results higher
for exponential service times.
Therefore, the analytic model underestimate the number of necessary
servers under a given stress condition. Drawing a horizontal
line in Figure~\ref{explogn}b, when we move in the direction
of increasing demand (that is, lowering $T_C$) we first cross
the blue (simulated) curves in all the considered cases.
This is an important fact to keep in mind if we want to use
the analytical model to find the best number of servers
in a particular operational scenario.
However, the discrepancies observed between the simulations
and the closed form expression for average MFPT are not significant
enough to cause a considerable effect in the estimation
of the optimal number of servers, given the separation among
the curves for different values of $L$.
Thus, for example, from the Figure~\ref{explogn}b
we can see that to obtain an average MFPT of 6 hs,
given $T_C = 14\,$min, 6 servers are needed, whereas
for $T_C = 10\,$, 8 servers are needed
under the same service conditions.

We also analyze the effect of asymmetry in the service
time distribution on average MFPT. For this purpose, we simulate
the average MFPT for a whole family of service time distribution
with fixed $T_S$ but changing the parameter {\tt sdlog} in the
interval $[0.25,1]$, where the minimum value corresponds to the
almost symmetric case and the maximum corresponds to the more
asymmetric situation.
The average MFPT as function of {\tt sdlog} parameter is shown
in Figure~\ref{lognorm}.
In all cases, the analytical prediction gives an acceptable
approximation for the simulated data.

\begin{figure}[!ht]
\begin{center}
\includegraphics[width=0.40\textwidth]{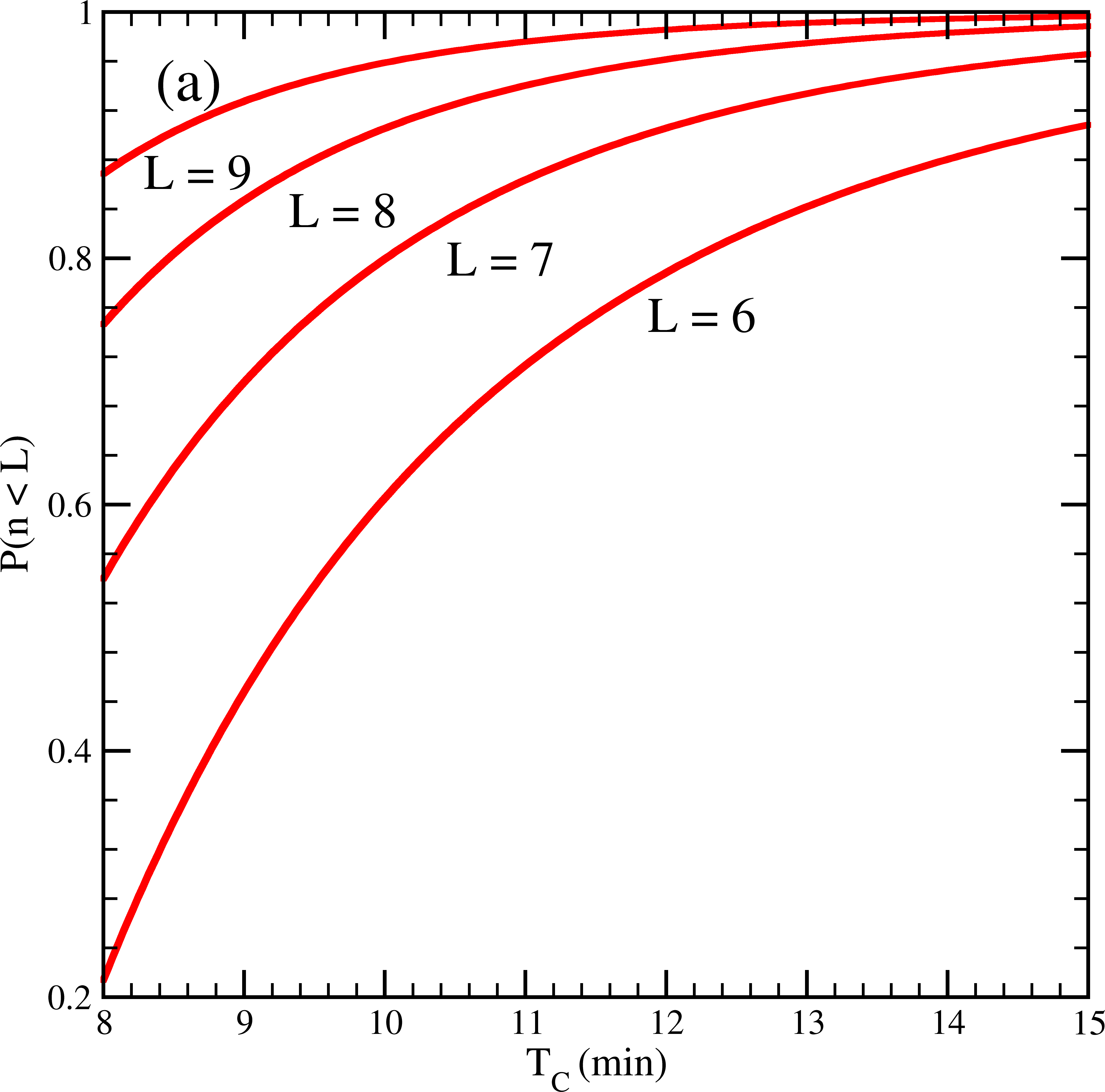}
\includegraphics[width=0.41\textwidth]{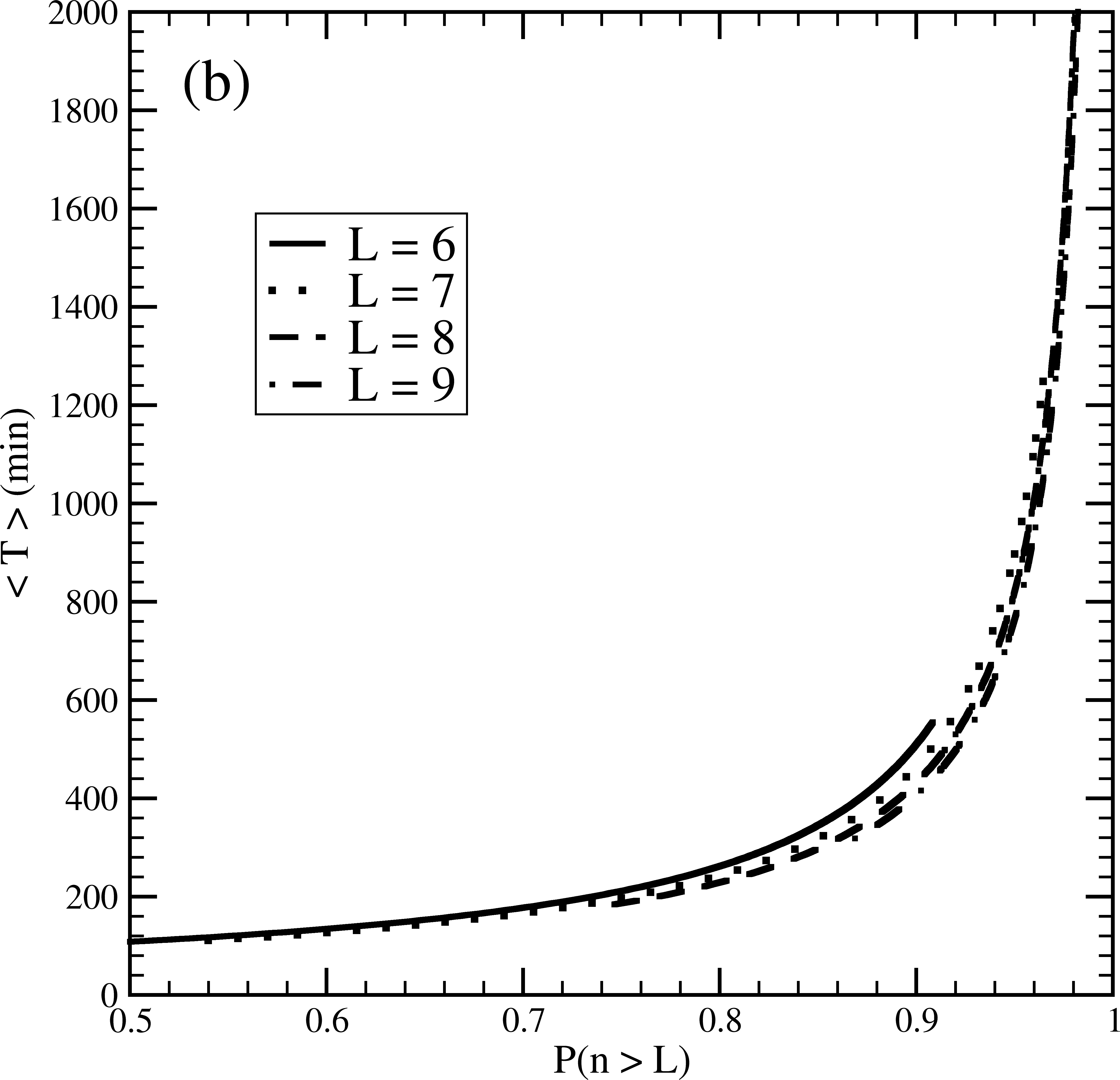}
\caption{{\bf (a)} Probability of one or more servers are free
vs $T_C$.
{\bf (b)} Average MFPT as function of the probability of servers
availability for $T_C \in [8,15]\,$min.
In both panels, $T_S=44.1\,$min.
}
\label{availability}
\end{center}
\end{figure}

Finally, in the left panel of Figure~\ref{availability} we plot
the probability that one or more servers are available
at the instant of a emergency call, $P(n < L)$~\cite{Hall71},
as function of the time between calls for a fixed mean service time.
We are using Eq.(\ref{Pn}) for the calculation of this probability.
In the right panel we plot the average MFPT as function of this
availability probability for the same values of $T_C$ given in
the left panel. We can observe a very strong sensitivity
of the average MFPT to the availability probability for high
values of system availability.
Thus, only controlling the availability of the system is not enough
to assure a long enough time to the critical condition.
A dispatcher observing a high availability probability value
might conclude that the system is running unreasonably idle;
however, the time to critical condition may be shorter than
the expectation.
More interesting, however, is that the function $<T>$ vs.\ $P(n < L)$
is practically independent of the number of servers.
%

\section{Concluding remarks}
\label{sec:fin}

MFPT is a useful KPI which allows estimate the running operative
lapse of a system, under a given stress condition, before
a service disruption.
In this work, we have presented a closed-form expression
to calculate the MFPT for a system of servers in parallel,
and we also provide a simulation framework for the MFPT.
Our formula, based on a birth--death process, only use
the average time between demands and the average service time.
Our results make it possible to predict the MFPT given the stress
of the system at a particular moment or to analyze the servers
shortage time under generic operating conditions by averaging over
the initial states of the system.
The main limitation of our results, as is often with analytical
exact results in queueing theory, is the assumption of exponential
distribution for service times.
The impact of this limiting assumption is confronted with results
of simulations using more realistic service distributions.
Our results indicate that our analytical formula is an acceptable
approximation under practical situations.
Interesting potential future work may be to consider
the implementation of accurate approximations for
the $M/G/L$ problem~\cite{Hok78} to the MFPT calculation.
In addition, our simulation scheme allows us to evaluate the
MFPT in any $GI/G/L/FCFS$ server configuration.


\vspace{0.2in}
\subsection*{Funding}
This work was partially supported by
{\em Sistema de Urgencias del Rosafe SA}, C{\'o}rdoba, Argentina
and by grant 33620180101258CB from
Se\-cre\-ta\-r\'\i a de Cien\-cia y Tec\-no\-lo\-g\'\i a de la
Uni\-ver\-si\-dad Na\-cio\-nal de C\'or\-doba, Argentina.

\subsection*{Conflicts of Interest}
The author declare no conflict of interest.
The funders had no role in the design of the study; in the collection,
analyses, or interpretation of data; in the writing of the manuscript,
or in the decision to publish the~results.

\subsection*{Abbreviations}
The following abbreviations are used in this manuscript:

\begin{tabular}{@{}ll}
DES  & Discrete-Event Simulator \\
EMS  & Emergency Medical Services \\
FCFS & First-Come First-Served \\
FPT  & First Passage-Time \\
KPI  & Key Performance Indicator \\
MFPT & Mean First-Passage Time \\
RNG  & pseudo-Random Number Generator \\
URG  & Sistema de Urgencias del Rosafe
\end{tabular}

\vspace{0.2in}
\section*{Appendices}
\appendix
\numberwithin{equation}{section}

\section{MFPT}
\label{sec:MFPT}

For a birth-death process with asymmetric and site dependent
transition probabilities ($w^+_k \neq w^-_k$),
the analytical expression for the MFPT with a reflecting
boundary condition at origin and an absorbing one at $L+1$
is given by
\begin{equation}
\begin{array}{lcl}
T(0) &=&
\displaystyle\sum_{k=0}^{L} \,\frac{1}{w^+_k} +
\displaystyle\sum_{k=0}^{L-1} \,\frac{1}{w^+_k} \,
\sum_{i=k+1}^{L} \prod_{j=k+1}^{i} \;\frac{w^-_j}{w^+_j} \;,
\\
T(1) &=& T(0) - \displaystyle\frac{1}{w^+_0} \;,
\\
T(n) &=& T(0) -
\displaystyle\sum_{k=0}^{n-1} \,\displaystyle\frac{1}{w^+_k} -
\displaystyle\sum_{k=0}^{n-2} \,\displaystyle\frac{1}{w^+_k} \,
\displaystyle\sum_{i=k+1}^{n-1} \prod_{j=k+1}^{i} \;\frac{w^-_j}{w^+_j}
\;\mbox{ for } 2 \leq n \leq L \,.
\label{MFPT:ra}
\end{array}
\end{equation}
For details and derivation of Eq.~(\ref{MFPT:ra}), see Section~6
in Ref.~\cite{PC03}.
In our model with $L$ servers, using Eq.~(\ref{Lservers})
and the parameter $\gamma$ defined in the text,
we can recast the products in Eq.~(\ref{MFPT:ra}) as
\begin{equation}
\prod_{j=k+1}^{i} \;\frac{w^-_j}{w^+_j} =
\gamma^{i-k} \prod_{j=k+1}^{i} j = \gamma^{i-k} \,\frac{i!}{k!} \,.
\label{prod1}
\end{equation}
Thus, we can also recast the sums in Eq.~(\ref{MFPT:ra}) as
\begin{equation}
\sum_{i=k+1}^{n-1} \prod_{j=k+1}^{i} \;\frac{w^-_j}{w^+_j} =
\frac{\gamma^{-k}}{k!} \sum_{i=k+1}^{n-1} i! \,\gamma^i \,,
\label{Sprod1}
\end{equation}
and
\begin{equation}
\sum_{k=0}^{n-2} \,\frac{1}{w^+_k} \,
\sum_{i=k+1}^{n-1} \prod_{j=k+1}^{i} \;\frac{w^-_j}{w^+_j} =
\frac{1}{\lambda} \sum_{k=0}^{n-2} \,
\frac{\gamma^{-k}}{k!} \sum_{i=k+1}^{n-1} i! \,\gamma^i \,.
\label{SSprod1}
\end{equation}
Replacing the last expression in Eq.~(\ref{MFPT:ra}),
we obtain in a direct manner Eq.~(\ref{MFPT:critical})
in Sec.~\ref{sec:model}.

\section{Steady state}
\label{sec:SS}

Following Ref.~\cite{Win03}, we can construct the steady state
for the problem of a Markov chain with a reflecting boundary
condition at the origin. In the long-run, the time-independent
probability of residence at state $n$, $\pi_n$, must satisfy
\begin{equation}
\begin{array}{rcl}
\omega^-_1 \,\pi_1 -\omega^+_0 \,\pi_0 &=& 0 \,,
\\
\rule{0cm}{0.5cm}
\omega^-_{n+1} \,\pi_{n+1} +\omega^+_{n-1} \,\pi_{n-1}
- (\omega^+_n + \omega^-_n) \,\pi_n &=& 0 \,,
\;\; \mbox{for $n \geq 1$} \,.
\end{array}
\label{stationary}
\end{equation}
Thus, we can proof by induction that
\begin{equation}
\pi_n = \frac{\omega^+_{n-1} \dots \omega^+_0}
           {\omega^-_n \dots \omega^-_1} \,\pi_0
= \prod_{j=1}^n \,\frac{\omega^+_{j-1}}{\omega^-_j} \,\pi_0 \,,
\;\mbox{for $n \geq 1$} \,.
\label{pin}
\end{equation}
From the normalization condition,
$\displaystyle\sum_{n=0}^{\infty} \pi_n = 1$,
results $\pi_0 = 1/S_{\pi}$, where
\begin{equation}
S_{\pi} = 1 + \sum_{n=1}^{\infty}
\,\prod_{j=1}^n \,\frac{\omega^+_{j-1}}{\omega^-_j} \,.
\label{Sdef}
\end{equation}
The existence of the steady state is then determined by the
convergence of the series in the Eq.~(\ref{Sdef}).

For the model given by Eq.~(\ref{Lservers}), the products
in Eq.~(\ref{pin}) and~(\ref{Sdef}) can be written as,
\begin{equation}
\prod_{j=1}^n \frac{\omega^+_{j-1}}{\omega^-_j}
= \left\{
\begin{array}{ll}
\displaystyle
\left( \frac{\lambda}{\mu} \right)^n \;\prod_{j=1}^n \,\frac{1}{j}
= \frac{\gamma^{-n}}{n!} \,,
& \mbox{for $0 \leq n \leq L$} \,,
\\
\rule{0cm}{1.0cm}
\displaystyle
\left( \frac{\lambda}{\mu} \right)^n
\;\prod_{j=1}^{L} \,\frac{1}{j} \,\prod_{j=L+1}^{n} \,\frac{1}{L}
= \frac{\gamma^{-n}}{L!} \,\frac{1}{L^{n-L}} \,,
& \mbox{for $n \geq L$} \,.
\end{array}
\right.
\label{prod}
\end{equation}
Substituting Eq.~(\ref{prod}) into Eq.~(\ref{Sdef}) we obtain,
\begin{equation}
S_{\pi} = \displaystyle
\sum_{n=0}^{L} \frac{\gamma^{-n}}{n!}
+ \frac{L^L}{L!} \,\sum_{n=L+1}^{\infty} (L \,\gamma)^{-n} \,.
\label{S1}
\end{equation}
The convergence of the series in the last expression only occurs
if $L \,\gamma > 1$. In this case,
\begin{equation}
\sum_{n=L+1}^{\infty} (L \,\gamma)^{-n}
= \frac{(L \,\gamma)^{-L}}{L \,\gamma - 1} \,.
\label{seriesconv}
\end{equation}
Substituting Eqs.~(\ref{prod}--\ref{seriesconv})
into Eq.~(\ref{pin}), we obtain
\begin{equation}
\pi_n = \displaystyle \frac{1}{S_{\pi}} \,\left\{
\begin{array}{ll}
\displaystyle \frac{\gamma^{-n}}{n!} \,,
& \mbox{for $0 \leq n \leq L$} \,, \\
\rule{0cm}{0.8cm}
\displaystyle \frac{L^L}{L!} \,(L \,\gamma)^{-n} \,,
& \mbox{for $n \geq L$} \,,
\end{array}
\right.
\label{pinfty}
\end{equation}
where
\begin{equation}
S_{\pi} = \displaystyle\sum_{n=0}^{L} \frac{\gamma^{-n}}{n!}
+ \displaystyle\frac{\gamma^{-L}}{L! \,(L \gamma-1)} \,.
\label{S}
\end{equation}
In this manner the long-term probability of having the system
{\em with calls in the waiting queue} results
\begin{equation}
\pi_q = \sum_{n=L+1}^{\infty} \pi_n
= \frac{\gamma^{-L}}{S_{\pi} \,L! \,(L \gamma-1)} \,.
\label{piq}
\end{equation}
The last expression is also is known as Erlang $C$-formula.

We now consider the case of a {\em finite} Markov chain
with reflecting boundaries at the origin and at site $L$.
The time-independent probabilities of residence, $P(n)$,
in the each state $n$ satisfy Eq.~(\ref{stationary}),
but supplemented with the additional reflecting condition at $L$,
\begin{equation}
\begin{array}{rcl}
\omega^-_1 \,P(1) -\omega^+_0 \,P(0) &=& 0 \,,
\\
\rule{0cm}{0.5cm}
\omega^-_{n+1} \,P(n+1) +\omega^+_{n-1} \,P(n-1)
- (\omega^+_n + \omega^-_n) \,P(n) &=& 0 \,,
\;\; \mbox{for $n \geq 1$} \,,
\\
\rule{0cm}{0.5cm}
\omega^+_{L-1} \,P(L-1) -\omega^-_L \,P(L) &=& 0 \,.
\end{array}
\label{eqPn}
\end{equation}
Following the above procedure, we find the first line of
Eq.~(\ref{prod}) again, which directly leads to
Eq.~(\ref{Pn}) in Sec.~\ref{sec:model}, where $S$
is the normalization on the interval $\left[ 0,L \right]$.


\vspace{0.2in}

\end{document}